\pdfoutput=1
\PassOptionsToPackage{bookmarks=false,colorlinks=true,linkcolor=blue,urlcolor=cyan,filecolor=magenta,citecolor=red,pdfstartview=FitV,hyperfootnotes=false,pdftitle={Quantum conformal Bondi-Metzner-Sachs field theories},pdfauthor={Daniel Grumiller, Iva Lovrekovic},pdfsubject={open null strings},pdfkeywords={open null strings and boundary carrollian conformal field theories},bookmarksopen=true}{hyperref}
\documentclass[aps,prl,reprint,superscriptaddress]{revtex4-2} 
  
\usepackage[dvipsnames]{xcolor}
\usepackage[bookmarks=false,colorlinks=true,linkcolor=blue,urlcolor=cyan,filecolor=magenta,citecolor=red,pdfstartview=FitV,hyperfootnotes=false,pdftitle={Quantum conformal Bondi-Metzner-Sachs field theories},pdfauthor={Daniel Grumiller, Iva Lovrekovic},pdfsubject={quantum conformal BMS},pdfkeywords={quantum conformal BMS algebra central extension flat space holography unitarity Kac-Moody level W-algebras},bookmarksopen=true]{hyperref}
\usepackage{amssymb,amsmath,amsfonts}
\usepackage{epsfig}
\usepackage{graphicx}
\usepackage{epstopdf}
\usepackage{braket}
\usepackage{dutchcal}
\usepackage{mathrsfs}
\usepackage{tikz}
\usepackage{pgfplots}
\pgfplotsset{compat=1.18}

\definecolor{CFTI}{HTML}{D73027}
\definecolor{CFTII}{HTML}{FC8D59}
\definecolor{CFTIII}{HTML}{1A9850}
\definecolor{CFTIV}{HTML}{4575B4}

\newcommand{\be}{\begin{eqnarray}}
	\newcommand{\ee}{\end{eqnarray}}
\newcommand{\beq}{\begin{eqnarray}}
	\newcommand{\eeq}{\end{eqnarray}}

\newcommand{\beqa}{\begin{eqnarray}}
	\newcommand{\eeqa}{\end{eqnarray}}

\newcommand{\tr}{\mathop{{\rm Tr}}}

\definecolor{gris}{rgb}{0.5,0.5,0.5}
\definecolor{darkgreen}{rgb}{0.0,0.5,0.0}

\DeclareMathOperator{\extdm}{d}
\newcommand{\extd}{\extdm \!}

\allowdisplaybreaks[0]

\begin{document}

		
\title{Unitary quantum conformal Bondi--Metzner--Sachs field theories}
	
\author{Daniel Grumiller} 
\email{grumil@hep.itp.tuwien.ac.at}
\affiliation{Institute for Theoretical Physics, TU Wien, Wiedner Hauptstrasse 8–10, 1040 Vienna, Austria}
\author{Iva Lovrekovic}
\email{iva.lovrekovic@tuwien.ac.at}
\affiliation{Institute for Theoretical Physics, TU Wien, Wiedner Hauptstrasse 8–10, 1040 Vienna, Austria}
	
\begin{abstract}
We impose unitarity constraints on two-dimensional field theories whose symmetries form the quantum conformal Bondi--Metzner--Sachs algebra, which combines conformal symmetries with asymptotic symmetries of asymptotically flat spacetimes. Positivity constraints put bounds on the central charge, $0<c<2.053$. Positivity of the associated vertex algebra allows only four central charges, $c=1,\,3/2,\,13/7,\,2$. Thus, only four candidates for unitary field theories remain. We calculate their spectra of primaries and show that they are compatible with BPS-like and holographic inequalities.
\end{abstract}
        
\maketitle



 
\section{Introduction}\label{sec:1}  

Relativistic quantum field theories are built on Poin\-car\'e symmetries: translations and (Lorentz\mbox{-)}ro\-ta\-tions. These symmetries have been deformed, contracted and extended in the past century. Arguably, their two most relevant extensions for phenomenology are conformal symmetries \cite{diFrancesco,Blumenhagen} and Bondi--Metzner--Sachs (BMS) symmetries \cite{Bondi:1962,Sachs:1962}. The former has dilatations and special conformal transformations (SCTs) as additional symmetries, thereby allowing to describe systems at criticality, like second order phase transitions in condensed matter \cite{diFrancesco,Cardy_1996}. The latter extends the translations to an infinite set of angle-dependent super-translations and has found applications in soft scattering, memory effects and flat space holography, see \cite{Bagchi:2012xr,Barnich:2012xq,Bagchi:2014iea,Strominger:2014pwa,Kapec:2016jld,Donnay:2022aba,Bagchi:2022emh,Bekaert:2024uuy} and  the reviews \cite{Donnay:2023mrd,Bagchi:2025vri,Ruzziconi:2026bix}.

Given these successes, it is perhaps not surprising that these extensions of Poincar\'e symmetries were ultimately combined into a single large algebra. Indeed, Barnich and Troessaert extended the rotations to an infinite set of super-rotations \cite{Barnich:2009se} (see also \cite{Campiglia:2014yka}), Donnay et al.~added super-dilatations \cite{Donnay:2020fof}, and Fuentealba et al.~completed the story in three dimensions (3d) by adding super-SCTs \cite{Fuentealba:2020zkf}. 

The classical limit of the algebra generating all these symmetries was called conformal BMS$_3$ algebra \cite{Fuentealba:2020zkf}. Its quantum version is also known as $W_{(2,2,2,1)}$ algebra. The subscripts refer to the conformal weights of the generators --- we shall see this explicitly when presenting the algebra \eqref{eq:1}-\eqref{eq:angelinajolie}. The associated field theory can be viewed as a two-dimensional (2d) warped conformal field theory (CFT) \cite{Detournay:2012pc,Hofman:2011zj} with additional symmetries. The physical spectra of such theories are determined by the charges $q$ and conformal weights $h$ of primaries.

The main goal of our Letter is to derive the spectra of all primaries for all unitary quantum conformal BMS$_3$ field theories and to discuss physical properties of these spectra, in particular, convexity conditions such as a BPS-type inequality or a holography inspired inequality. The key results are summarized in Figs.~\ref{fig:1}, \ref{fig:2} and Table \ref{tab:2}. The key take away lesson is that there are four admissible quantum conformal BMS$_3$ field theories, the spectra of which we construct in this Letter.


\section{Quantum conformal BMS$_3$ algebra}

Denoting super-rotations by $L_n$ (with $n\in\mathbb{Z}$), super-dilatations by $J_n$, super-translations by $M_n$, super-SCTs by $K_n$, the Virasoro central charge by $c$, and the $\hat u(1)_k$ Kac--Moody level by $k$, the quantum conformal BMS$_3$ algebra is given by the commutation relations \cite{Fasquel:2020iqf}
\begin{align}
[L_n,\,L_m] &= (n-m)\,L_{n+m} + \frac{c}{12}\,\big(n^3-n\big)\,\delta_{n+m,\,0} \label{eq:1} \\
[L_n,\,J_m] &= -m\,J_{n+m} \label{eq:2} \\
[L_n,\,M_m] &= (n-m)\,M_{n+m} \label{eq:3} \\
[L_n,\,K_m] &= (n-m)\,K_{n+m} \label{eq:4} \\
[J_n,\,J_m] &= k\,n\,\delta_{n+m,\,0} \label{eq:5} \\
[J_n,\,M_m] &= M_{n+m} \label{eq:6} \\
[J_n,\,K_m] &= -K_{n+m} \label{eq:7} \\
[M_n,\,M_m] &= [K_n,\,K_m] = 0 \label{eq:8} \\
[M_n,\,K_m] &= \frac{c}{12}\,\big(n^3-n\big)\delta_{n+m,\,0} + (n-m)\,L_{n+m} \nonumber \\ 
&\!\!\!\!\!\!\!\!\!\!\!\!\!\!\!\!\!\!\!\!\!\!\!\!\!\!\!\! + \frac{c}{12k}\big(n^2+m^2-nm-1\big)J_{n+m} + \frac{1-\frac{c}{4k}}{c-1}(n-m)\Lambda^{(2)}_{n+m} 
\nonumber \\
&\quad + \frac{c\,(4k-1)}{2k^2\,(c-1)}\,\Lambda_{n+m}^{(3,p)} + \frac{c}{3k^2}\,\Lambda_{n+m}^{(3,q)}\,.
\label{eq:9}
\end{align}

The last commutator contains the composite operators
($r,s\in\mathbb{Z}$ in the sums)~\footnote{%
The quantities $\Lambda^{(2)}$ and $\Lambda^{(3,p)}$ are defined such that they are Virasoro primaries of respective weights $2$ and $3$ --- which explains the precise values of the relative coefficients --- while $\Lambda^{(3,q)}$ is a quasiprimary of weight $3$; hence the labels $p$ and $q$.
}
\begin{align}
\Lambda^{(2)}_n &= L_n - \frac{c}{2k}\,\sum_r\colon\!J_{n-r}\,J_r \colon \label{eq:10} \\
\Lambda^{(3,p)}_n &= \sum_r\colon\!L_{n-r}\,J_r\colon - \frac{c+2}{6k}\sum_{r,s}\colon\!J_{n-r-s}\,J_r\,J_s\colon \label{eq:11} \\
\Lambda^{(3,q)} &= \sum_{r,s}\colon\!J_{n-r-s}\,J_r\,J_s\colon \label{eq:12}
\end{align}
with normal ordering defined in the usual way, i.e., 
\begin{align}
    \colon\!\mathcal{O}_{n-r}\,J_r \colon  & = \begin{cases}
  \mathcal{O}_{n-r}\,J_r \;\textrm{if}\; r\geq 0 \\
  J_r\,\mathcal{O}_{n-r} \;\textrm{if}\; r<0 
\end{cases} \\
    \colon\!\mathcal{O}_{n-r-s}\,J_r\,J_s\colon &= \begin{cases}
  \colon\!\mathcal{O}_{n-r-s}\,J_r\colon\,J_s \;\textrm{if}\; s\geq 0 \\
  J_s\,\colon\!\mathcal{O}_{n-r-s}\,J_r\colon \;\textrm{if}\; s<0 
\end{cases}
\end{align}
for any operator $\mathcal{O}_n\in\{L_n,J_n,M_n,K_n\}$.

\subsection{Central charge and level duality}

A remarkable aspect of the quantum conformal BMS$_3$ algebra \eqref{eq:1}-\eqref{eq:12} is that the Virasoro central charge $c$ cannot be independent from the level $k$ as a consequence of the Jacobi identities \cite{Fasquel:2020iqf,Yu:2022bcp,Gupta:2023fmp}. It is straightforward to verify that all the Jacobi identities hold for the algebra \eqref{eq:1}-\eqref{eq:12} except for the ones that involve two super-translation and a super-SCT~\footnote{%
Also the Jacobi identities involving two super-SCTs and one super-translation do not hold without imposing \eqref{eq:angelinajolie}. However, they do not provide any new information due to the involutive automorphism of the algebra \eqref{eq:1}-\eqref{eq:12}, $J_n\to-J_n$, $M_n\leftrightarrow K_n$. Since the literature on this subject has a somewhat confusing history, we present key aspects of the Jacobi identities in the Supplemental Material (SM).
}. The latter imply
\begin{equation}
c = -12k + 28 - \frac{30}{k+1}\,.
\label{eq:angelinajolie}
\end{equation}
In other words, the quantum conformal BMS$_3$ algebra is only consistent if the relation \eqref{eq:angelinajolie} holds. From now on, we always will assume this to be the case.

The algebra \eqref{eq:1}-\eqref{eq:angelinajolie} is also known as $W_{(2,2,2,1)}$ algebra and was first presented explicitly by Fasquel \cite{Fasquel:2020iqf}~\footnote{%
Fasquel's paper builds on well-known work by Drinfeld and Sokolov \cite{Drinfeld:1984qv}, Feigin and Frenkel \cite{Feigin:1990pn}, and others, e.g., \cite{Bais:1990bs,Bowcock:1991zk,deBoer:1992sy,Kac:2015haq}, see \cite{Bouwknegt:1992wg,deBoer:1995cqx,Blumenhagen} for physics perspectives on $W$-algebras. In essence, Fasquel performed a Drinfeld--Sokolov reduction of $\mathfrak{\widehat{sp}}_{k-2}(4)$ to $W_{(2,2,2,1)}$. The classical limit of this algebra appeared for the first time in a physics context and was dubbed ``conformal BMS$_3$ algebra'' but also referred to as $W_{(2,2,2,1)}$ algebra by Fuentealba et al.~\cite{Fuentealba:2020zkf}, see also \cite{Yu:2022bcp,Gupta:2023fmp}.
}. 

An immediate consequence of the key relation \eqref{eq:angelinajolie} is that not all real values of $c$ are allowed for real $k$. The central charge is either bigger than $40+12\sqrt{10}\approx77.95$ or smaller than $40-12\sqrt{10}\approx2.053$. Moreover, at the special point $k=\tfrac14$, the coefficients in \eqref{eq:9} are finite by inserting \eqref{eq:angelinajolie}, e.g., $\lim_{k\to1/4}(4k-1)/(c-1)=5/9$.

More strikingly, for every level $k$ there is a dual level, $\tilde{k}=(3-2k)/[2(k+1)]$, that yields the same value for the central charge $c$ as $k$. 

This implies, in particular, a dual semiclassical regime (in the sense that $c\to+\infty$) not considered so far besides the more obvious one, $k\to-\infty$ discussed already in \cite{Fuentealba:2020zkf}. This dual semiclassical regime is obtained when $k$ tends to $-1$ from below, $k=\lim_{\epsilon\to0^+}-1-\epsilon$.

\subsection{Notable subalgebras}

Since the quantum conformal BMS$_3$ algebra \eqref{eq:1}-\eqref{eq:angelinajolie} generates numerous symmetries, it has many well-studied subalgebras. We mention here the most notable ones.

The subalgebra generated by super-rotations is a chiral half of the 2d conformal algebra, see, e.g., \cite{Balasubramanian:2009bg}. Thus, quantum conformal BMS$_3$ field theories could be regarded as chiral CFTs with extended symmetries.

The subalgebra generated by super-rotations $L_n$ and super-translations $M_n$ is BMS$_3$ \cite{Ashtekar:1996cd}, though not with the usual central extension expected from general relativity \cite{Barnich:2006av} but instead with the one implied by conformal gravity \cite{Bagchi:2012yk}. It is equivalent to the 2d Carrollian conformal algebra \cite{Bagchi:2010zz,Duval:2014uva}. See \cite{Bagchi:2025vri,Ruzziconi:2026bix} for recent reviews.

The subalgebra generated by super-rotations $L_n$ and super-dilatations $J_n$ is the warped conformal algebra \cite{Detournay:2012pc,Hofman:2011zj}. Thus, quantum conformal BMS$_3$ field theories are warped CFTs with extended symmetries; as we shall see in Eq.~\eqref{eq:42} below, the states are labeled by the same pair of quantum numbers as in warped CFTs.

This abundance of symmetries may lead us to expect unitarity to be very constraining, and indeed it will be.


\section{Restrictions from unitarity}

Up to now, we have not considered restrictions imposed by unitarity. The simplest restrictions are non-negativity of the central charge $c\geq0$ and the $\hat u(1)_k$ level $k\geq0$. Since in the quantum conformal BMS$_3$ algebra both are related by \eqref{eq:angelinajolie} we get lower and upper bounds on $k$, both coming from positivity of $c$:
\begin{equation} 
0.1396\approx\frac23-\sqrt{\frac{5}{18}}\leq k\leq\frac23+\sqrt{\frac{5}{18}}\approx 1.194 
\label{eq:17}
\end{equation}

In the allowed interval \eqref{eq:17} of the level $k$, the central charge varies between $c=0$ at the ends of the interval and the maximal value
\begin{equation}
    c_\ast=40-12\sqrt{10}\approx 2.053 \quad\;\, k_\ast=\sqrt{\frac{5}{2}}-1\approx 0.581\,.
    \label{eq:18}
\end{equation}

Thus, already at this stage we can conclude that a unitary CFT$_2$ with quantum conformal BMS$_3$ symmetry can only have a central charge in the interval $c\in[0,\,2.053)$ and a level in the interval $k\in(0.1396,\,1.194)$. In particular, semiclassical values (large $c$) are excluded. This is not surprising but well-known from studies of higher spin theories in 3d that contain a $\hat u(1)_k$ current algebra: at large $k$ and large $c$, the signs of level and central charge are opposite \cite{Castro:2012bc}~\footnote{%
There are exceptions to this general guideline as shown for the Feigin--Semikhatov algebra \cite{Afshar:2012hc} but they require a large $N$ limit where $N$ is the rank of the gauge group in the underlying CS description, so these exceptions are irrelevant for the present work.
}.

While the necessary condition \eqref{eq:17} is already quite constraining, on general grounds we expect further constraints from unitarity that restrict $k$ and $c$ to certain rational values. Fasquel proved that only four admissible levels are compatible with positivity of the underlying vertex algebra and hence can be unitary \cite{Fasquel:2020iqf}. We list them in the conventions of our Letter:
\begin{equation}
\textrm{necessary\;for\;unitarity:}\quad k\in\Big\{\frac14,\,\frac13,\,\frac34,\,\frac23\Big\} 
\label{eq:19}
\end{equation} 
This implies that there are at most four unitary CFTs that have quantum conformal BMS$_3$ symmetries, which we label by Roman numerals.

\setlength{\tabcolsep}{6pt} 
\begin{table}[h!tb]
\begin{center}
\begin{tabular}{|l|l|l|l|l|l|l|}\hline
& $c$ & $k$ & $N_{\textrm{\tiny irre}}$ & $N_{\textrm{\tiny prim}}$  & $h_{\textrm{\tiny high}}$ & $h_{\textrm{\tiny low}}-\frac{c}{24}$\\\hline
CFT I & $1$ & $\frac14$ & 4 & 5 & $\frac12$ & $\frac{1}{12}$ \\
CFT II & $\frac32$ & $\frac13$ & 9 & 12 & $\frac23$ & $\frac{1}{24}$ \\
CFT III & $\frac{13}{7}$ & $\frac34$ & 12 & 21 & $\frac32$ & $\frac{1}{84}$ \\
CFT IV & $2$ & $\frac23$ & 18 & 30 & $\frac43$ & $0$ \\\hline
\end{tabular}
\caption{%
Four admissible theories and their spectral data. $c$: central charge. $k$: $\hat u(1)_k$ level.  $N_{\textrm{\tiny irre}}$: number of irreducible $W_{(2,2,2,1)}$ modules. $N_{\textrm{\tiny prim}}$: number of primaries. $h_{\textrm{\tiny high}}$: conformal weight of highest primaries. $h_{\textrm{\tiny low}}$: conformal weight of lowest non-vacuum primaries}
\label{tab:2}
\end{center}
\end{table}

We summarize the values of $k$ and the associated central charges $c$ in Table \ref{tab:2} (ordered by ascending values of $c$), together with additional information about the spectrum of primaries that we discuss in the next sections.


\section{Physical states and primaries}

 We label physical states by their charge $q$ and their conformal weight $h$, defined as eigenvalues of $J_0$ and $L_0$.
\begin{equation}
J_0\,|\psi\rangle = q\,|\psi\rangle\qquad\qquad L_0\,|\psi\rangle = h\,|\psi\rangle
\label{eq:42}
\end{equation}
We denote primary states as $|q,\,h\rangle$. They are defined by the usual highest weight conditions ($n>0$)
\begin{equation}
L_n|q,\,h\rangle = J_n|q,\,h\rangle = M_n|q,\,h\rangle = K_n|q,\,h\rangle= 0
\end{equation}
The vacuum, $|0,\,0\rangle$, is a primary. As usual \cite{diFrancesco,Blumenhagen,Cardy_1996}, descendant states are obtained by acting with creation operators $L_{-n},\,J_{-n},\,M_{-n},\,K_{-n}$ on primary states.


\subsection{Spectral flow}

The algebra \eqref{eq:1}-\eqref{eq:angelinajolie} has three useful automorphisms: rescaling, $M_n\to\lambda{M}_n$, $K_n\to{K}_n/\lambda$, charge conjugation $J_n\to-J_n$, $M_n\leftrightarrow{K}_n$, and spectral flow \cite{Schwimmer:1986mf,Fasquel:2020iqf}
\begin{align}
\psi(L_n) &= L_n - J_n + \frac{k}{2}\,\delta_{n,\,0} & \psi(J_n) &= J_n - k\,\delta_{n,0} \nonumber \\
\psi(M_n) &= M_{n-1} & \psi(K_n) &= K_{n+1} \,.
\label{eq:sf}
\end{align}

When we act with spectral flow \eqref{eq:sf} on the module $L(q,\,h)$ we generate a new module with different weights,
\begin{equation}
    \psi\big(L(q,\,h)\big) = L(q+i-1-k,\,h-q-i+1+\tfrac k2)
    \label{eq:whatever}
\end{equation}
where $i\geq 1$ is the dimension of the top space of $L(q,\,h)$, i.e., the number of primary states in the module~\footnote{%
The module $L(q,\,h)$ contains at least one primary and all their descendants and is labeled by the weights of a primary $|q,\,h\rangle$ that additionally obeys $K_0|q,\,h\rangle=0$. A module $L(q,\,h)$ can contain several primaries with the same weights $h$ but with charges that differ by integers. 
}. The top space is spanned by primary states $M_0^j|q,\,h\rangle$ with $j=0,\dots,i-1$, which for $i\geq2$ increases the value of the charge by an integer,
\begin{equation}
J_0\,M_0^j\,|q,\,h\rangle = (q+j)\,M_0^j\,|q,\,h\rangle\,.
\label{eq:whynolabel}
\end{equation}
The dimension of the top space $i$ is determined from the conditions $M_0^i|q,\,h\rangle=0$ and $M_0^{i-1}|q,\,h\rangle\neq0$. Fasquel's Proposition 6.2 \cite{Fasquel:2020iqf} shows that $i$ has to be a positive integer given by $i_+,\,i_-$ or $i_++i_-$, with $i_\pm=\tfrac12-q\pm\sqrt{\tfrac14+k(k-1)-q(q-1)+2h(k+1)}$~\footnote{%
To derive $i_\pm$, start with a primary that obeys $K_0|q,\,h\rangle=0$, act on it with $K_0M_0$, insert the expression \eqref{eq:9}, and obtain $K_0M_0|q,\,h\rangle\propto{g}(q,\,h)|q,\,h\rangle$ with $g(q,\,h)=q[k(k-1)+2h(k+1)-2q^2]$. By Proposition 6.2, the quantity $h(q,\,h)=-\tfrac2i\,\sum_{m=0}^{i-1}g(q+m,\,h)=(i+2q-1)[i^2+i(2q-1)+k(1-k)-2h(1+k)+2q(q-1)]$ must vanish when $i$ is the dimension of the top space. This leads to the three possibilities $i=i_+$, $i=i_-$, or $i=i_++i_-$.   
}. If more than one are positive integers then $i$ is given by the lowest one. In particular, for $q=0$ we always get $i=i_++i_-=1$.

The vacuum module $L(0,\,0)$ also has $i=1$. Acting on it up to two times with spectral flow yields
\begin{equation}
 L(0,\,0) \xrightarrow{\psi} L(-k,\,\tfrac k2) \xrightarrow{\psi}  L(-2k,\,2k) \xrightarrow{\psi}\, ?
\label{eq:lalapetz}
\end{equation}
The second relation holds because the dimension $i$ of the top space of $L(-k,\,\tfrac{k}{2})$ is always $1$ since either $i_
+=1$ or $i_-=1$. What happens when we spectrally flow away from $L(-2k,\,2k)$ depends on the value of $k$, since we get $i_+=k$ and $i_-=1+3k$. From these expressions and the knowledge that one of them (or their sum) has to be a positive integer we see that $k$ can only take certain rational values, $p/3$ or $p/4$ with $p\in\mathbb{Z}^+$, cf.~Eq.~\eqref{eq:19}.


\subsection{Spectrum of primaries}

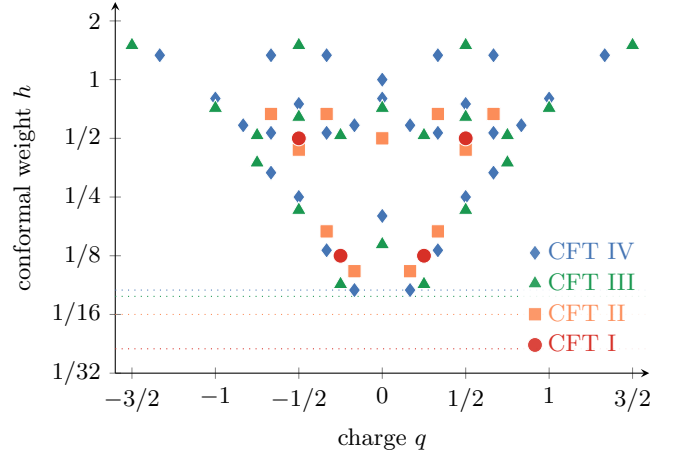
\begin{figure}[h!tb]
\begin{center}
\begin{tikzpicture}
\begin{axis}[
    width=\linewidth,
    height=0.75*\linewidth,
    xmin=-1.6, xmax=1.6,
    ymin=1/32, ymax=2.5,
    ymode=log,
    log basis y=2,
    xlabel={charge $q$},
    ylabel={conformal weight $h$},
    axis lines=left,
    grid=none,
    xtick={-3/2,-1,-1/2,0,1/2,1,3/2},
    xticklabels={$-3/2$,$-1$,$-1/2$,$0$,$1/2$,$1$,$3/2$},
    ytick={1/32,1/16,1/8,1/4,1/2,1,2},
    yticklabels={$1/32$,$1/16$,$1/8$,$1/4$,$1/2$,$1$,$2$},
    tick label style={font=\small},
    label style={font=\small},legend style={
        font=\small,
        at={(1.0,0.2)},
        anchor=east,
        draw=none,
        fill=white,
        fill opacity=0.9,
        text opacity=1,
        /tikz/column 2/.style={column sep=7pt},
    }, 
    legend cell align=left,
    mark options={draw=white, line width=0.3pt},
    clip=false,
]
\addplot[CFTI,thin,dotted,forget plot] coordinates {(-1.6,1/24) (1.6,1/24)};
\addplot[CFTII,thin,dotted,forget plot] coordinates {(-1.6,1/16) (1.6,1/16)};
\addplot[CFTIII,thin,dotted,forget plot] coordinates {(-1.6,13/168) (1.6,13/168)};
\addplot[CFTIV,thin,dotted,forget plot] coordinates {(-1.6,1/12) (1.6,1/12)};
\addplot[
    only marks,
    mark=diamond*,
    mark size=3.0pt,
    color=CFTIV,
    fill=CFTIV,
] coordinates {
    (2/3,1/3)
    (1/3,2/15)
    (0,1/5)
    (1/6,1/12)
    (-1/6,1/12)
    (0,1)
    (-1/3,2/15)
    (-2/3,1/3)
    (-1/2,1/4)
    (1/2,1/4)
    (-1/3,8/15)
    (2/3,8/15)
    (-1/6,7/12)
    (5/6,7/12)
    (-1/2,3/4)
    (1/2,3/4)
    (-5/6,7/12)
    (1/6,7/12)
    (-2/3,8/15)
    (1/3,8/15)
    (-1,4/5)
    (0,4/5)
    (1,4/5)
    (-4/3,4/3)
    (-1/3,4/3)
    (2/3,4/3)
    (-2/3,4/3)
    (1/3,4/3)
    (4/3,4/3)
};
\addlegendentry[CFTIV]{CFT IV}
\addplot[
    only marks,
    mark=triangle*,
    mark size=3.2pt,
    color=CFTIII,
    fill=CFTIII,
] coordinates {
    (3/4,3/8)
    (0,1/7)
    (1/4,5/56)
    (-1/4,5/56)
    (-3/4,3/8)
    (-1/2,3/14)
    (1/2,3/14)
    (-1/4,29/56)
    (3/4,29/56)
    (-1/2,9/14)
    (1/2,9/14)
    (-3/4,29/56)
    (1/4,29/56)
    (-1,5/7)
    (0,5/7)
    (1,5/7)
    (-3/2,3/2)
    (-1/2,3/2)
    (1/2,3/2)
    (3/2,3/2)
};
\addlegendentry[CFTIII]{CFT III}
\addplot[
    only marks,
    mark=square*,
    mark size=2.5pt,
    color=CFTII,
    fill=CFTII,
] coordinates {
    (1/3,1/6)
    (1/6,5/48)
    (0,1/2)
    (-1/6,5/48)
    (-1/3,1/6)
    (-1/2,7/16)
    (1/2,7/16)
    (-2/3,2/3)
    (1/3,2/3)
    (-1/3,2/3)
    (2/3,2/3)
};
\addlegendentry[CFTII]{CFT II}
\addplot[
    only marks,
    mark=*,
    mark size=2.8pt,
    color=CFTI,
    fill=CFTI,
] coordinates {
    (1/4,1/8)
    (-1/4,1/8)
    (-1/2,1/2)
    (1/2,1/2)
};
\addlegendentry[CFTI]{CFT I}
\end{axis}
\end{tikzpicture}
\caption{Spectra of non-vacuum primaries for all four CFTs. Dotted lines are at $\frac{c}{24}$ of corresponding CFT to visualize gaps.
All spectra respect charge conjugation, $|q,\,h\rangle\to|-q,\,h\rangle$.}
\label{fig:1}
\end{center}
\end{figure}

We exploit now spectral flow starting from the vacuum module to determine the spectrum of all non-vacuum primaries for CFT I, see the four circles in Fig.~\ref{fig:1}.

For CFT I we have $k=\tfrac14$. Thus, the two non-vacuum primaries achieved by spectrally flowing from the vacuum \eqref{eq:lalapetz} are given by $|-\tfrac14,\,\tfrac18\rangle$ and $|-\tfrac12,\,\tfrac12\rangle$. The dimension of the top space of the module $L(-\tfrac12,\,\tfrac12)$ is given by $i_++i_-=2$, so that spectral flow \eqref{eq:whatever} yields the primary $|\tfrac14,\,\tfrac18\rangle$. Spectrally flowing from the associated module $L(\tfrac14,\,\tfrac18)$ reproduces the vacuum module. It turns out there are no additional modules, i.e., for CFT I all of them are generated by spectrally flowing from the vacuum. The complete list of non-vacuum primaries is
\begin{equation}
\textrm{CFT\,I:}\quad|\pm\tfrac14,\,\tfrac18\rangle\qquad\qquad |\pm\tfrac12,\,\tfrac12\rangle\,.
\label{eq:CFTI}
\end{equation}
Note that the primaries $|\pm\tfrac12,\,\tfrac12\rangle$ belong to the same module $L(-\tfrac12,\,\tfrac12)$, which has $i=2$ as top space dimension. Consistently, the values of the charges of the two primaries in this module differ by $1$.

For CFT II most of the work was done already in \cite{Fasquel:2020iqf} and we can read off the weights of the modules from their Example 7.6. Translating this into weights of non-vacuum primaries yields the complete list
\begin{align}
\textrm{CFT\,II:}&&|\pm\tfrac16,\,\tfrac{5}{48}\rangle &&|\pm\tfrac13,\,\tfrac16\rangle &&|\pm\tfrac12,\,\tfrac{7}{16}\rangle \nonumber\\
&&|0,\,\tfrac12\rangle &&|\pm\tfrac23,\,\tfrac23\rangle &&|\pm\tfrac13,\,\tfrac23\rangle 
\label{eq:CFTII}
\end{align}

We have applied the same methods to CFT III and IV (details can be found in the SM) and find, respectively, 20 and 29 non-vacuum primaries for these CFTs. This establishes the data points displayed in Table \ref{tab:2} and Figs.~\ref{fig:1} and \ref{fig:2}. Thus, we have succeeded in determining the full spectra of primaries for all four quantum conformal BMS$_3$ field theories~\footnote{%
In \cite{Fasquel:2020iqf}, necessary conditions for unitarity, in particular positivity of the underlying vertex algebra, were established, and unitarity of CFT I--IV was conjectured in Conjecture 7.7. Unitarity was subsequently proven for CFT I--III in \cite{Fasquel_2025}, see their Theorem 7.18. For CFT IV, the question of unitarity remains unresolved. We nevertheless include CFT IV in our discussion. For brevity, in the following we use ``unitarity'' to refer to the established unitarity of CFT I--III and, for CFT IV, to the necessary conditions for unitarity established in \cite{Fasquel:2020iqf}.
}.


\section{BPS bound}

Unitarity of the $\hat u(1)_k$ sector yields a BPS-type inequality that puts an upper bound on the charge $q$ in terms of the conformal weight $h$, see, e.g., \cite{Detournay:2012pc}~\footnote{%
BPS bounds are well-known from supersymmetric theories \cite{Witten:1978mh,Weinberg:2000cr}. Even though we have no supersymmetry, the algebra $W_{(2,2,2,1)}$ has supersymmetry-like features, with $M_n$ and $K_n$ playing the role of the supersymmetry generators and $J_n$ playing the role of the R-symmetry. 
}. 
\begin{equation}
\textrm{BPS:}\quad h \geq \frac{q^2}{2k}
\label{eq:BPS}
\end{equation}
The inequality \eqref{eq:BPS} holds for all four CFTs and is saturated for four non-vacuum primaries in each CFT.  We call non-vacuum primaries that saturate \eqref{eq:BPS} extremal. All non-vacuum primaries in CFT I are extremal.  

The structure of extremal primaries is the same for all four CFTs: there are two charge-conjugated highest-weight primaries $|\pm 2k,\,2k\rangle$ and two charge-conjugated primaries with half the charge $|\pm k,\,k/2\rangle$. This is a consequence of spectral flow from the vacuum, which generates $|-k,\,\tfrac{k}{2}\rangle$ and $|-2k,\,2k\rangle$, while charge conjugation gives the corresponding positive-charge primaries, see \eqref{eq:lalapetz}. Thus, all extremal primaries lie on the vacuum orbit. 

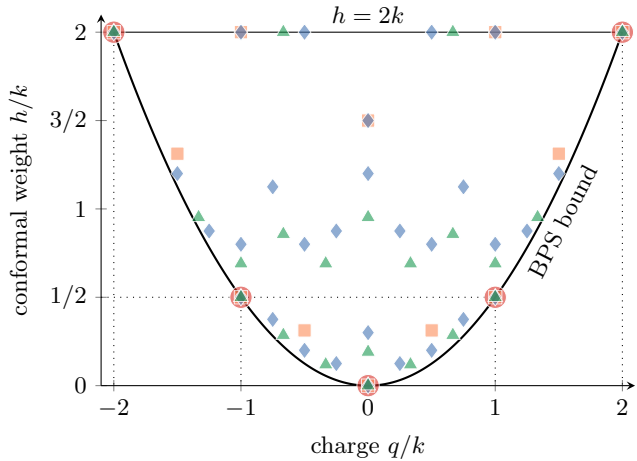
\begin{figure}[h!tb]
\begin{tikzpicture}
\begin{axis}[width=\linewidth, height=0.75*\linewidth, xmin=-2.1,  xmax=2.1, ymin=0, ymax=2.1, xlabel={charge $q/k$}, ylabel={conformal weight $h/k$}, axis lines=left, grid=none, xtick={-2,-1,0,1,2}, xticklabels={$-2$,$-1$,$0$,$1$,$2$}, ytick={0,1/2,1,3/2,2}, yticklabels={$0$,$1/2$,$1$,$3/2$,$2$}, tick label style={font=\small}, mark options={draw=white, line width=0.3pt}, clip=false]
\addplot[domain=-2:2, samples=200, thick,black, forget plot] {x^2/2}node[pos=0.8, yshift=-12, sloped, above, font=\small, fill=white, fill opacity=0.85, text opacity=1, inner sep=1.5pt] {BPS bound};
\addplot[black,thin,forget plot] coordinates {(-2,2) (2,2)} node[pos=0.5, above] {$h=2k$};
\addplot[black,thin,dotted,forget plot] coordinates{(-2.1,1/2) (1,1/2)};
\addplot[black,thin,dotted,forget plot] coordinates{(-1,1/2) (-1,0)};
\addplot[black,thin,dotted,forget plot] coordinates{(1,1/2) (1,0)};
\addplot[black,thin,dotted,forget plot] coordinates{(-2,2) (-2,0)};
\addplot[black,thin,dotted,forget plot] coordinates{(2,2) (2,0)};
\addplot[
only marks,
mark=*,
mark size=4.2pt,
color=CFTI,
fill=CFTI,
fill opacity=0.6,
draw opacity=0.8,
] coordinates {
(0,0)
(1,1/2)
(-1,1/2)
(-2,2)
(2,2)
};
\addplot[
only marks,
mark=square*,
mark size=2.5pt,
color=CFTII,
fill=CFTII,
fill opacity=0.6,
draw opacity=0.8,
] coordinates {
(0,0)
(1,1/2)
(1/2,5/16)
(0,3/2)
(-1/2,5/16)
(-1,1/2)
(-3/2,21/16)
(3/2,21/16)
(-2,2)
(1,2)
(-1,2)
(2,2)
};
\addplot[
only marks,
mark=diamond*,
mark size=3.2pt,
color=CFTIV,
fill=CFTIV,
fill opacity=0.6,
draw opacity=0.8,
] coordinates {
(0,0)
(1,1/2)
(1/2,1/5)
(0,3/10)
(1/4,1/8)
(-1/4,1/8)
(0,3/2)
(-1/2,1/5)
(-1,1/2)
(-3/4,3/8)
(3/4,3/8)
(-1/2,4/5)
(1,4/5)
(-1/4,7/8)
(5/4,7/8)
(-3/4,9/8)
(3/4,9/8)
(-5/4,7/8)
(1/4,7/8)
(-1,4/5)
(1/2,4/5)
(-3/2,6/5)
(0,6/5)
(3/2,6/5)
(-2,2)
(-1/2,2)
(1,2)
(-1,2)
(1/2,2)
(2,2)
};
\addplot[
only marks,
mark=triangle*,
mark size=3.2pt,
color=CFTIII,
fill=CFTIII,
fill opacity=0.6,
draw opacity=0.8,
] coordinates {
(0,0)
(1,1/2)
(0,4/21)
(1/3,5/42)
(-1/3,5/42)
(-1,1/2)
(-2/3,2/7)
(2/3,2/7)
(-1/3,29/42)
(1,29/42)
(-2/3,6/7)
(2/3,6/7)
(-1,29/42)
(1/3,29/42)
(-4/3,20/21)
(0,20/21)
(4/3,20/21)
(-2,2)
(-2/3,2)
(2/3,2)
(2,2)
};
\end{axis}
\end{tikzpicture}
\caption{Uniformized spectra, BPS bound, and upper bound. Dotted lines aid visually to find extremal primaries.}
\label{fig:2}
\end{figure}

In Fig.~\ref{fig:2} we plot uniformized versions of the spectra where both the charge and the conformal weight are rescaled by $k$ so that the BPS bound \eqref{eq:BPS} is visible as a universal parabola, $h/k=(q/k)^2/2$. The extremal primaries of different CFTs lie on top of each other. 

Finally, there is an upper bound on the conformal weight, $h\leq2k$. It arises because the only modules saturating this bound have either charge labels $q=-2k$ (for all CFTs) or $q=-k$ (for CFT II and IV). The dimension of the associated top spaces are always $i=1+4k$ for CFT I, III and $i=1+3k$ for CFT II, IV. Applying spectral flow \eqref{eq:whatever} to such modules always produces values of $h<2k$. Thus, if we are at the top in Fig.~\ref{fig:2} spectral flow pushes us back down. Moreover, all the modules strictly below the bound obey the stronger inequality  $h\leq\frac{3k}{2}$. The addition to $h$ induced by spectral flow \eqref{eq:whatever} can therefore never exceed the bound $h\leq2k$. 


\section{Gap in spectra}

The non-vacuum primary spectra of CFT I-IV are gapped and obey  the inequality
\begin{equation}
\textrm{gap:}\quad    h \geq \frac{c}{24}\,.
    \label{eq:inequality}
\end{equation}
Moreover, the inequality \eqref{eq:inequality} is strict for CFT I-III and saturates for the two primaries $(\pm\frac16,\,\frac{1}{12})$ of CFT IV. 

We have visualized this in Fig.~\ref{fig:1} by drawing horizontal lines at the values $\frac{c}{24}$ so that it is evident that the spectra lie above these lines and the size of the gap is visible.

We derive now the values for the level $k$ for which the bound \eqref{eq:inequality} holds, using the following working assumption: the lowest-lying non-vacuum primaries have a charge $q_0$ that is not (half-)integer. Note that this assumption is true for CFT I and III ($q_0=\tfrac14$) and also for CFT II and IV ($q_0=\tfrac16$). Since $q_0$ is not (half-)integer, the sum $i_++i_-=1-2q_0$ [see the definitions after \eqref{eq:whynolabel}] cannot be integer.

We additionally assume that even for non-unitary theories either $i_+=1$ or $i_-=1$ for the lowest-lying non-vacuum primary (again, this is true for CFT I-IV). We then solve $i_+=1$ (or $i_-=1$) for the weight of the lowest-lying primary, $h_0=[q_0^2-k(k-1)/2]/(k+1)$, and plug this result into the left-hand side of \eqref{eq:inequality}. On the right hand side, we insert the relation \eqref{eq:angelinajolie}. Together, they establish an inequality for the level,
\begin{equation}
-1< k \leq 6q_0^2 + \frac12\,.
    \label{eq:condition}
\end{equation}    
For $q_0=\tfrac16$ (the value for CFT II and IV) we get $k\leq\tfrac23$, which saturates for CFT IV. Thus, our derivation of the inequality \eqref{eq:condition} explains why CFT IV saturates the  bound \eqref{eq:inequality}, and why CFT II obeys it. For $q_0=\tfrac14$ (the value for CFT I and III) we get $k\leq\tfrac78$, which explains why CFT I and III strictly obey the bound \eqref{eq:inequality}. Finally, for $-1<k\leq\tfrac12$ the value of $q_0$ is irrelevant (as long as our initial assumptions apply) and the bound \eqref{eq:inequality} always holds~\footnote{%
Curiously, in the semiclassical limit $k\to-\infty$ the BPS state $|-|k|,\,\tfrac{|k|}{2}\rangle$ always marginally violates the bound, $\tfrac{c}{24}=\tfrac{|k|}{2}+\tfrac76+\mathcal{O}(1/|k|)>\tfrac{|k|}{2}=h$.
}.

\subsection{Holographic interpretation of gap}

The inequality \eqref{eq:inequality} is a purely field theoretic result. However, it is significant in a holographic context, see, e.g., the discussion in \cite{Castro:2011zq}. From a gravity perspective, the inequality \eqref{eq:inequality} is natural and captures the gap in the spectrum of classical gravity solutions in 3d with negative cosmological constant \cite{Banados:1992wn,Banados:1992gq}.

While it is a bit of a stretch to impose a gravity interpretation on a model where the central charge is of order unity, CFT I-IV are similar in this regard to the critical and tricritical Ising model discussed from a holographic perspective in \cite{Castro:2011zq}, which are the only minimal model CFTs that obey the inequality \eqref{eq:inequality} for all primaries.





\section{Outlook}


We conclude with a discussion of open issues, applications, and generalizations.

Whether CFT IV --- the only one that saturates the holographic bound \eqref{eq:inequality}  --- is unitary remains unresolved and is a key open question.

For CFT I–III, and for CFT IV if it proves to be unitary, the next logical step is to write down partition functions using results for the characters in \cite{Fasquel:2020iqf}. One can then check physical and mathematical properties of these partition functions, including thermal behavior, modular properties, or the existence of Schwarzian-type sectors, see \cite{Ghosh:2019rcj} for such discussions in a CFT context and \cite{Aggarwal:2025hji} in a Carrollian CFT context. Given that quantum conformal BMS$_3$ field theories feature Carrollian and warped CFT symmetries, it could be rewarding to investigate the corresponding generalization of \cite{Aggarwal:2025hkb}.

There are several natural algebraic extensions. A straightforward one is to include supersymmetry. Classically, this was already done in \cite{Fuentealba:2020zkf}, while a quantum treatment could reveal whether the unitarity restrictions and spectral bounds found here persist or acquire supersymmetric refinements. Similarly, one could consider further symmetry enhancements such as those studied in \cite{Fuentealba:2024thk}. It would also be interesting to investigate whether adding sufficiently many higher-spin generators can evade the semiclassical unitarity obstruction, analogous to the mechanism found for $W_N^{(2)}$ theories \cite{Afshar:2012nk}, where increasing the rank allows arbitrarily large central charges while maintaining unitarity.

A more ambitious generalization is a conformal version of BMS$_4$, where the 2d super-rotation algebra becomes two Virasoro algebras and super-translations carry two integer labels, see \cite{Barnich:2009se} for details. The freedom to modify the brackets between super-rotations and super-translations through boundary conditions \cite{Grumiller:2019fmp} may provide useful input for constructing a quantum conformal BMS$_4$ algebra.

Finally, our results invite a holographic interpretation. The classical asymptotic symmetry analysis \cite{Fuentealba:2020zkf} of the $\mathfrak{so}(3,2)$ CS theory \cite{Horne:1988jf} related to conformal gravity \cite{Deser:1982vy} suggests a natural candidate bulk dual for quantum conformal BMS$_3$ field theories, while the relation between the bulk CS level and the boundary $\hat u(1)_k$ level remains to be understood. The semiclassical identification $k_{\textrm{\tiny CS}}=-k$ \cite{Fuentealba:2020zkf} cannot hold literally at the quantum level if the properly normalized CS level $k_{\textrm{\tiny CS}}$ is integer, since none of the $k$ allowed by unitarity are integer, see Eq.~\eqref{eq:19}. The relation between the bulk CS level $k_{\textrm{\tiny CS}}$ and the boundary $\hat u(1)_k$ level therefore requires quantum and/or normalization corrections. A particularly interesting question is whether the holographic gap \eqref{eq:inequality}, including its saturation in CFT IV, has a natural explanation in the bulk description.





\paragraph{}\paragraph{}\acknowledgments 

We thank Florian Ecker for discussions and Bin Chen and Zhe-Fei Yu for correspondence.

This work was supported by the Austrian Science Fund (FWF) [Grants DOI: \href{https://www.fwf.ac.at/en/research-radar/10.55776/P36619}{10.55776/P36619}, \href{https://www.fwf.ac.at/en/research-radar/10.55776/PAT1871425}{10.55776/PAT1871425}, and \href{https://www.fwf.ac.at/forschungsradar/10.55776/V1052}{10.55776/V1052}].  


%


\clearpage
\onecolumngrid
\setcounter{equation}{0}
\setcounter{table}{0}
\renewcommand{\theequation}{S\arabic{equation}}
\renewcommand{\thetable}{S\arabic{table}}

\section{Supplemental Material}

In the Supplemental Material, we provide details on the derivation of the relation between the central charge $c$ and the level $k$ for the quantum conformal Bondi--Metzner--Sachs (BMS) algebra, also known as $W_{(2,2,2,1)}$ algebra. Where applicable, our results agree with those in the literature, with the exception of the relation $c=-12k+4$ presented in v1--v4 of Ref.~\cite{Yu:2022bcp}. We have communicated with the authors, who agree with the relation \eqref{eq:angelinajolie} given in the main text. To facilitate the comparison, we start with a dictionary of notations and conventions. 

In the final three subsections, we list all weights of all primaries for CFT I-IV defined in the main text, discuss and list all spectral flow orbits, and give details on the Chern--Simons (CS) formulation of conformal gravity.

\subsection{Dictionary of notations and conventions}\label{app:A}

Since the notations and conventions in previous work \cite{Fasquel:2020iqf,Fuentealba:2020zkf,Yu:2022bcp,Gupta:2023fmp} differ, we present a dictionary in Table \ref{tab:1}. Our notations and conventions agree with \cite{Yu:2022bcp} and are highlighted in bold faced letters in Table \ref{tab:1}. 

\setlength{\tabcolsep}{6pt} 
\begin{table}[h!bt]
\begin{center}
\begin{tabular}{|l|l|l|l|l|l|l|}
\hline
gravity entity & field theory entity & weights & Ref.~\cite{Fasquel:2020iqf}& Ref.~\cite{Fuentealba:2020zkf} & \textbf{Ref.~\cite{Yu:2022bcp}} & Ref.~\cite{Gupta:2023fmp} \\\hline
super-rotations & Virasoro generators & $(0,2)$ & $L_n$ & $J_n$ & $\boldsymbol{L_n}$ & $L_n$ \\
super-dilatations & $\hat u(1)_k$ Kac--Moody generators & $(0,1)$ & $J_n$ & $-iD_n$ & $\boldsymbol{J_n}$ & $H_n$ \\
super-translations & $+$ charged weight-2 primary & $(1,2)$ & $G^+_n$ & $P_n$ & $\boldsymbol{M_n}$ & $P^0_n-P_n^1$ \\
super-SCTs & $-$ charged weight-2 primary & $(-1,2)$ & $-\frac{c}{6(k+2)^2(k+1)}\,G^-_n$ & $-\frac12\,K_n$ & $\boldsymbol{K_n}$ & $P_n^0+P_n^1$ \\
gravitational coupling & Virasoro central charge & n.~A. & $c$ & $\frac{c}{12}$ & $\boldsymbol{c}$ & $c$ \\
CS level & Kac--Moody level & n.~A. & $k+2$ & $-\tilde c=-k$ & $\boldsymbol{k}$ & $\kappa$\\
\hline
\end{tabular}
\caption{Dictionary of notations and conventions}
\label{tab:1}
\end{center}
\end{table}

The first two columns provide, respectively, the gravitational and field theoretic interpretation of the corresponding entity. The third column provides the charge (viz., the $J_0$-eigenvalue) and the conformal weight (viz., the $L_0$-eigenvalue) of that entity (n.~A.~means `not applicable'). The last four columns translate notations and conventions between previous work; the ones used in our work coincide with the bold faced column. 

The rescaling of $G^-$ in row and column four shows that we need to be careful for $k=0$ and $k=1$ (here we mean again our $k$). We disregard the former case since we assume non-vanishing level, $k\neq 0$. The latter case corresponds to $c=k=1$ and leads to a well-defined algebra in the conventions of \cite{Fasquel:2020iqf} but to a spurious singularity in \cite{Yu:2022bcp} and thus also in the present work. However, since this value for $k$ is incompatible with unitarity we do not have a strong incentive to choose a different normalization for the super-SCTs. Thus, we exclude $k=\pm1,0$ in the present work.

\subsection{Ansatz for quantum conformal BMS$_3$ algebra}

To derive the quantum BMS$_3$ algebra we keep all linear commutators from the main text \eqref{eq:1}-\eqref{eq:8} and replace the last commutator by the ansatz
\begin{multline}
[M_n,\,K_m] = \frac{c}{12}\,\big(n^3-n\big)\,\delta_{n+m,\,0} + a_1\,(n-m)\,L_{n+m} + a_2\,\big(n^2+m^2-nm-1\big)\,J_{n+m} \\+ a_3\,(n-m)\,\Lambda^{(2)}_{n+m} + a_4\,\Lambda_{n+m}^{(3,p)} + a_5\,\Lambda_{n+m}^{(3,q)} \label{eq:9a}
\end{multline}
where $a_i$ are numerical coefficients that we fix by imposing the Jacobi identities; they may depend on $k$ and $c$ but not on $n$ or $m$. 

This ansatz is motivated by the following considerations: For large values of the level $k$ we must recover the classical conformal BMS$_3$ algebra of \cite{Fuentealba:2020zkf}, which fixes the commutators \eqref{eq:1}-\eqref{eq:8} of the main text. These commutators are invariant under arbitrary (finite) rescalings of the generators $M_n$ and $K_n$. We have fixed (part of) this redundancy by selecting the coefficient in the anomalous term on the right hand side of \eqref{eq:9a} to be $\frac{c}{12}$, exactly as in \cite{Yu:2022bcp}. Finally, we wrote down all the terms that could appear on the right hand side of \eqref{eq:9a} with arbitrary coefficients that will receive quantum corrections from normal ordering compared to their classical values. The latter can be read off from \cite{Fuentealba:2020zkf} (using the dictionary in Table \ref{tab:1}).

\subsection{Jacobi identities}\label{app:C}

We impose now all the Jacobi identities $\{G_1,G_2,G_3\}=[[G_1,G_2],G_3]+[[G_2,G_3],G_1]+[[G_3,G_1],G_2]=0$, where $G_i$ comprise all the generators $L_n,J_n,M_n,K_n$. All Jacobi identities hold already, except for the following four: $\{M_n,K_m,L_l\}=0$, $\{M_n,K_m,J_l\}=0$, $\{M_n,K_m,M_l\}=0$, and $\{M_n,K_m,K_l\}=0$. The last one follows trivially from the penultimate by virtue of the involutive automorphism $M_n\leftrightarrow K_n$, $J_n\to -J_n$ of the algebra \eqref{eq:1}-\eqref{eq:12}, so we have to verify only three remaining Jacobi identities.

For checking the Jacobi identities it is useful to collect formulas of commutators involving composite operators. The first commutator in each line below reveals the (quasi-)primary behavior of the corresponding operator. In the last line, we define $\colon\!J^2\colon\!_n=\sum_r\colon\!J_{n-r}J_r\colon$ and use analogous definitions for $\colon\!MJ\colon\!_n$ and $\colon\!LJ\colon\!_n$ further below.
\begin{align}
[L_n,\,\Lambda^{(2)}_m] &= (n-m)\,\Lambda^{(2)}_{n+m} & [J_n,\,\Lambda^{(2)}_m] &= (1-c)\,n\,J_{n+m}
\label{eq:app1} \\
[L_n,\,\Lambda^{(3,p)}_m] &= (2n-m)\,\Lambda^{(3,p)}_{n+m} & [J_n,\,\Lambda^{(3,p)}_m] &= -\frac c2\,n\,\Lambda^{(2)}_{n+m}
\label{eq:app2} \\
[L_n,\,\Lambda^{(3,q)}_m] &= (2n-m)\,\Lambda^{(3,q)}_{n+m} + \frac{k}{2}\,\big(n^3-n\big)\,J_{n+m} & [J_n,\,\Lambda^{(3,q)}_m] &= 3k\,n\,\colon\!J^2\colon\!_{n+m}\,.
\label{eq:app3}
\end{align}

Applying the commutation relations above, the identities $\{M_n,K_m,L_l\}=0$ imply
\begin{equation}
a_1 = 1\qquad\qquad a_5 = \frac{4}{k}\,a_2\,.
\label{eq:13}
\end{equation}
The identities $\{M_n,K_m,J_l\}=0$ fix the remaining coefficients, yielding
\begin{equation}
a_2 =\frac{c}{12k} \qquad\qquad  a_3 = \frac{1-\frac{c}{4k}}{c-1} \qquad\qquad a_4 = \frac{c\,(4k-1)}{2k^2\,(c-1)} \,.
\label{eq:14}
\end{equation}
Inserting these coefficients into \eqref{eq:9a} establishes \eqref{eq:9} in the main text.

The Jacobi identities $\{M_n,K_m,M_l\}=0$ reduce to $[[M_n,K_m],M_l]=[[M_l,K_m],M_n]$, i.e., we only need the part of $[[M_n,K_m],M_l]$ that is antisymmetric in $n$ and $l$. By virtue of the commutation relations 
\begin{align}
    &(n-m)\,[\colon\!J^2\colon\!_{n+m},\,M_l] - (l-m)\,[\colon\!J^2\colon\!_{l+m},\,M_n] = 2(n-l)\,\colon\!MJ\colon\!_{n+m+l}+(n-l)\,(n+l+1)\,M_{n+m+l}\label{eq:app4}\\
    &[\colon\!LJ\colon\!_{n+m},\,M_l] - [\colon\!LJ\colon\!_{l+m},\,M_n] = 2(n-l)\,\colon\!MJ\colon\!_{n+m+l}+(n-l)\,(m+1)\,M_{n+m+l}\label{eq:app5}\\
    &[\colon\!J^3\colon\!_{n+m},\,M_l] -  [\colon\!J^3\colon\!_{l+m},\,M_n] = 3(n-l)\,\colon\!MJ\colon\!_{n+m+l} + \frac12\,(n-l)\,(n+l+2m+3)\,M_{n+m+l}
    \label{eq:app6}
\end{align}
we obtain
\begin{multline}
0=[[M_n,K_m],M_l]-[[M_l,K_m],M_n]=(n-l)\,\Big[\big(2N_{JJ}+2N_{LJ}+3N_{JJJ}\big)\,\big(\colon\!MJ\colon\!_{n+m+l}+\frac12\,(m+1)\,M_{n+m+l}\big)\\
+(n+l-m)\,\Big(1+\frac{c}{12k}+N_{JJ}\,\big(1-\frac{2k}{c}\big)+\frac12\,N_{JJJ}\Big)\,M_{n+m+l}\Big]
\label{eq:app7}
\end{multline}
with the coefficients determined from \eqref{eq:9a}-\eqref{eq:14},
\begin{equation}
N_{JJ} = -\frac{c\,a_3}{2k} = \frac{c\,(4k-c)}{8k^2\,(c-1)}\quad\qquad N_{LJ} = a_4 = \frac{c\,(4k-1)}{2k^2\,(c-1)} \qquad\quad N_{JJJ} = a_5 -\frac{(c+2)\,a_4}{6k} = \frac{c\,(c-12k+2)}{12k^3\,(c-1)}\,.
\label{eq:app8}
\end{equation} 

To cancel the first line in the Jacobiator \eqref{eq:app7} we need to impose the condition 
\begin{equation}
2N_{JJ}+2N_{LJ}+3N_{JJJ}=\frac{c\,\big(c\,(k+1)+12k^2-16k+2\big)}{4k^3\,(c-1)}=0 
\label{eq:app9}
\end{equation}
which is trivially solved by $c=0$ (a case we discard as physically irrelevant) and non-trivially solved by the relation \eqref{eq:angelinajolie} between central charge $c$ and level $k$. The special cases $k=\frac14$ (or $k=1$) and $c=1$ are compatible with the Jacobi identities, despite the singular factor $c-1$ in the denominator of \eqref{eq:app9}.

The second line in the Jacobiator \eqref{eq:app7} has a coefficient that evaluates to
\begin{equation}
1+\frac{c}{12k}+N_{JJ}\,\Big(1-\frac{2k}{c}\Big)+\frac12\,N_{JJJ} = \frac{c\,(2k+1)\,\big(c\,(k+1)+12k^2-16k+2\big)}{24k^3\,(c-1)}
\end{equation}
and vanishes if the relation \eqref{eq:angelinajolie} in the main text holds. Thus, we only get one constraint from the Jacobi identities $\{M_n,K_m,M_l\}=0$ (or, equivalently, $\{M_n,K_m,K_l\}=0$), namely the key relation \eqref{eq:angelinajolie} in the main text.

\subsection{List of weights of all primaries for CFT I-IV}\label{app:E}

We list here the weights $(q,\,h)$, where $q$ is the charge and $h$ the conformal weight, of all primaries for CFT I-IV of the main text, using the methods of \cite{Fasquel:2020iqf}. The dimension of the top space of the module $L(q,\,h)$ is denoted by $i$.

\paragraph{\color{CFTI}{CFT I.}} Key data: $c=1$, $k=\frac14$, 4 irreducible $W_{(2,2,2,1)}$ modules from Proposition 6.5 (b) of \cite{Fasquel:2020iqf}: 
\begin{align}
i&=j=1: && \big(0,\,0\big) && \big(\tfrac14,\,\tfrac18\big)\\
i&=1,\,j=2: && \big(-\tfrac14,\,\tfrac18\big) && \\
i&=2,\,j=1: && \big(-\tfrac12,\,\tfrac12\big) &&
\label{eq:21}
\end{align}
Thus, there are $3*1+1*2=5$ primaries with the following weights.
\begin{align}
&& \color{CFTI}{\big|0,\,0\big>} && \color{CFTI}{\big|\pm\tfrac14,\,\tfrac18\big>} && \color{CFTI}{\big|\pm\tfrac12,\,\tfrac12\big>} 
\end{align}

\paragraph{\color{CFTII}{CFT II.}} Key data: $c=\frac32$, $k=\frac13$, 9 irreducible $W_{(2,2,2,1)}$ modules from Proposition 6.5 (a) of \cite{Fasquel:2020iqf}: 
\begin{align}
i&=j=1: && \big(0,\,0\big) && \big(\tfrac13,\,\tfrac16\big) && \big(\tfrac16,\,\tfrac{5}{48}\big)\\
i&=1,\,j=2: && \big(0,\,\tfrac12\big) && \big(-\tfrac16,\,\tfrac{5}{48}\big) && \big(-\tfrac13,\,\tfrac16\big) \\
i&=2,\,j=1: && \big(-\tfrac12,\,\tfrac{7}{16}\big) && \big(-\tfrac23,\,\tfrac23\big) && \big(-\tfrac13,\,\tfrac23\big)
\label{eq:20}
\end{align}
Thus, there are $6*1+3*2=12$ primaries with the following weights.
\begin{align}
&& \color{CFTII}{\big|0,\,0\big>} && \color{CFTII}
{\big|\pm\tfrac16,\,\tfrac{5}{48}\big>} && \color{CFTII}
{\big|\pm\tfrac13,\,\tfrac16\big>} && \color{CFTII}
{\big|\pm\tfrac12,\,\tfrac{7}{16}\big>} && \color{CFTII}
{\big|0,\,\tfrac12\big>} && \color{CFTII}{\big|\pm\tfrac23,\,\tfrac23\big>} && \color{CFTII}{\big|\pm\tfrac13,\,\tfrac23\big>}
\end{align}

\paragraph{\color{CFTIII}{CFT III.}} Key data: $c=\frac{13}{7}$, $k=\frac34$, 12 irreducible $W_{(2,2,2,1)}$ modules from Proposition 6.5 (b) of \cite{Fasquel:2020iqf}: 
\begin{align}
i&=j=1: && \big(0,\,0\big) && \big(\tfrac34,\,\tfrac38\big)\\
i&=1,\,j=2: && \big(0,\,\tfrac17\big) && \big(\tfrac14,\,\tfrac{5}{56}\big) \\
i&=1,\,j=3: && \big(-\tfrac14,\,\tfrac{5}{56}\big) && \\
i&=1,\,j=4: && \big(-\tfrac34,\,\tfrac{3}{8}\big) && \\
i&=2,\,j=1: && \big(-\tfrac12,\,\tfrac{3}{14}\big) && \big(-\tfrac14,\,\tfrac{29}{56}\big) \\
i&=2,\,j=2: && \big(-\tfrac12,\,\tfrac{9}{14}\big) && \big(-\tfrac34,\,\tfrac{29}{56}\big) \\
i&=3,\,j=1: && \big(-1,\,\tfrac57\big) && \\
i&=4,\,j=1: && \big(-\tfrac32,\,\tfrac32\big) &&
\label{eq:22}
\end{align}
Thus, there are $6*1+4*2+1*3+1*4=21$ primaries with the following weights.
\begin{align}
&& \color{CFTIII}{\big|0,\,0\big>}&& \color{CFTIII}{\big|\pm \tfrac14,\,\tfrac{5}{56}\big>} && \color{CFTIII}{\big|0,\,\tfrac17\big>}  && \color{CFTIII}{\big|\pm\tfrac12,\,\tfrac{3}{14}\big>} && \color{CFTIII}{\big|\pm\tfrac34,\,\tfrac38\big>}  && \color{CFTIII}{\big|\pm\tfrac34,\,\tfrac{29}{56}\big>} \\
&& \color{CFTIII}{\big|\pm\tfrac14,\,\tfrac{29}{56}\big>} 
&& \color{CFTIII}{\big|\pm\tfrac12,\,\tfrac{9}{14}\big>} && \color{CFTIII}{\big|\pm1,\,\tfrac57\big>} && \color{CFTIII}{\big|0,\,\tfrac57\big>} && \color{CFTIII}{\big|\pm\tfrac32,\,\tfrac32\big>} && \color{CFTIII}{\big|\pm\tfrac12,\,\tfrac32\big>} 
\end{align}

\paragraph{\color{CFTIV}{CFT IV.}}  Key data: $c=2$, $k=\frac23$, 18 irreducible $W_{(2,2,2,1)}$ modules from Proposition 6.5 (a) of \cite{Fasquel:2020iqf}: 
\begin{align}
i&=j=1: && \big(0,\,0\big) && \big(\tfrac23,\,\tfrac13\big) && \big(\tfrac13,\,\tfrac{2}{15}\big)\\
i&=1,\,j=2: && \big(0,\,\tfrac15\big) && \big(\tfrac16,\,\tfrac{1}{12}\big) && \big(-\tfrac16,\,\tfrac{1}{12}\big) \\
i&=1,\,j=3: && \big(0,\,1\big) && \big(-\tfrac13,\,\tfrac{2}{15}\big) && \big(-\tfrac23,\,\tfrac13\big) \\
i&=2,\,j=1: && \big(-\tfrac12,\,\tfrac14\big) && \big(-\tfrac13,\,\tfrac{8}{15}\big) && \big(-\tfrac16,\,\tfrac{7}{12}\big) \\
i&=2,\,j=2: && \big(-\tfrac12,\,\tfrac34\big) && \big(-\tfrac56,\,\tfrac{7}{12}\big) && \big(-\tfrac23,\,\tfrac{8}{15}\big) \\
i&=3,\,j=1: && \big(-1,\,\tfrac45\big) && \big(-\tfrac43,\,\tfrac43\big) && \big(-\tfrac23,\,\tfrac43\big) 
\label{eq:28}
\end{align}
Thus, there are $9*1+6*2+3*3=30$ primaries with the following weights.
\begin{align}
&& \color{CFTIV}{\big|0,\,0\big>} && \color{CFTIV}{\big|\pm\tfrac16,\,\tfrac{1}{12}\big>} && \color{CFTIV}{\big|\pm\tfrac13,\,\tfrac{2}{15}\big>} && \color{CFTIV}{\big|0,\,\tfrac15\big>}&& \color{CFTIV}{\big|\pm\tfrac12,\,\tfrac14\big>} && \color{CFTIV}{\big|\pm\tfrac23,\,\tfrac13\big>}  && \color{CFTIV}{\big|\pm\tfrac23,\,\tfrac{8}{15}\big>}  && \color{CFTIV}{\big|\pm\tfrac13,\,\tfrac{8}{15}\big>} && \color{CFTIV}{\big|\pm\tfrac56,\,\tfrac{7}{12}\big>} \\ && \color{CFTIV}{\big|\pm\tfrac16,\,\tfrac{7}{12}\big>}   
&& \color{CFTIV}{\big|\pm\tfrac12,\,\tfrac34\big>} && \color{CFTIV}{\big|\pm1,\,\tfrac45\big>} && \color{CFTIV}{\big|0,\,\tfrac45\big>} && \color{CFTIV}{\big|0,\,1\big>}&& \color{CFTIV}{\big|\pm\tfrac43,\,\tfrac43\big>} && \color{CFTIV}{\big|\pm\tfrac23,\,\tfrac43\big>}  && \color{CFTIV}{\big|\pm\tfrac13,\,\tfrac43\big>} &&
\end{align}

Note that CFT I, II, III, and IV have, respectively, one, two, three, and four neutral primaries, $(q=0,\,h\geq 0)$.

\subsection{List of spectral flow orbits}\label{app:G}

All modules of all CFTs considered here can be organized into spectral flow orbits. We list here all these orbits for the four admissible CFTs.

\paragraph{\color{CFTI}{CFT I.}} Starting from the vacuum module $L(0,\,0)$ we generate all modules of CFT I by spectral flow \eqref{eq:whatever}:
\begin{equation}
\textrm{Orbit\;I}_1: \qquad 
\color{CFTI}{L(0,\,0)\;\xrightarrow{\psi}\;L(-\tfrac14,\,\tfrac18)\;\xrightarrow{\psi}\;L(-\tfrac12,\,\tfrac12)\;\xrightarrow{\psi}\;L(\tfrac14,\,\tfrac18)\;\xrightarrow{\psi}\;L(0,\,0)}
    \label{eq:sf1}
\end{equation}
A distinguishing feature of CFT I is that it has only the vacuum orbit I$_1$ of spectral flow.

\paragraph{\color{CFTII}{CFT II.}} Besides the vacuum orbit there is the lowest-weight orbit that we generate starting from a module with the lowest conformal weight above zero, $L(-\tfrac16,\,\tfrac{5}{48})$. Before writing down both orbits we explain the precise value of the weight $\tfrac{5}{48}$. As mentioned after \eqref{eq:whynolabel}, either $i_\pm$ or their sum must be integer. However, for $q=-\tfrac16$ the sum, $i_++i_-=\tfrac43$ is not integer and hence one of $i_\pm$ must be. Since the lowest-weight modules always have a one-dimensional top space, this integer must be $1$. This establishes the relation
\begin{equation}
i_+=1\quad\textrm{or}\quad i_-=1:\qquad h = \frac{q^2-\frac12\,k(k-1)}{k+1}
    \label{eq:sf2}
\end{equation}
between the weight $h$ and the charge $q$ of the lowest-weight state. Plugging $q=-\tfrac16$ and $k=\tfrac13$ into
\eqref{eq:sf2} yields $h=\tfrac{5}{48}$. The two spectral flow orbits of CFT II are given by
\begin{align}
&\textrm{Orbit\;II}_1: \qquad 
\color{CFTII}{L(0,\,0)\;\xrightarrow{\psi}\;L(-\tfrac13,\,\tfrac16)\;\xrightarrow{\psi}\;L(-\tfrac23,\,\tfrac23)\;\xrightarrow{\psi}\;L(0,\,\tfrac12)\;\xrightarrow{\psi}\;L(-\tfrac13,\,\tfrac23)\;\xrightarrow{\psi}\;L(\tfrac13,\,\tfrac16)\;\xrightarrow{\psi}\;L(0,\,0)} \\
&\textrm{Orbit\;II}_2: \qquad 
\color{CFTII}{L(-\tfrac16,\,\tfrac{5}{48})\;\xrightarrow{\psi}\;L(-\tfrac12,\,\tfrac{7}{16})\;\xrightarrow{\psi}\;L(\tfrac16,\,\tfrac{5}{48})\;\xrightarrow{\psi}\;L(-\tfrac16,\,\tfrac{5}{48})} 
\label{eq:sf3}
\end{align}
A distinguishing feature of CFT II is that it has only the vacuum orbit II$_1$ and the lowest-weight orbit II$_2$ of spectral flow.

\paragraph{\color{CFTIII}{CFT III.}} Besides the vacuum orbit starting from $L(0,\,0)$ and the lowest-weight orbit starting from $L(-\tfrac14,\,\tfrac{5}{56})$ --- again we use \eqref{eq:sf2} to determine $h$ from $q=-\tfrac14$ and $k=\tfrac34$ --- there is another orbit of spectral flow. We can generate it starting from a module with the same charge as the lowest-weight module but assuming a 2d top space
\begin{equation}
i_+=2\quad\textrm{or}\quad i_-=2:\qquad h = \frac{q^2+q+1-\frac12\,k(k-1)}{k+1}
    \label{eq:sf4}
\end{equation}
which yields $L(-\tfrac14,\,\tfrac{29}{56})$. The three spectral flow orbits of CFT III are given by
\begin{align}
&\textrm{Orbit\;III}_1: \qquad 
\color{CFTIII}{L(0,\,0)\;\xrightarrow{\psi}\;L(-\tfrac34,\,\tfrac38)\;\xrightarrow{\psi}\;L(-\tfrac32,\,\tfrac32)\;\xrightarrow{\psi}\;L(\tfrac34,\,\tfrac38)\;\xrightarrow{\psi}\;L(0,\,0)} \\
&\textrm{Orbit\;III}_2: \qquad 
\color{CFTIII}{L(-\tfrac14,\,\tfrac{5}{56})\;\xrightarrow{\psi}\;L(-1,\,\tfrac{5}{7})\;\xrightarrow{\psi}\;L(\tfrac14,\,\tfrac{5}{56})\;\xrightarrow{\psi}\;L(-\tfrac12,\,\tfrac{3}{14})\;\xrightarrow{\psi}\;L(-\tfrac14,\,\tfrac{5}{56})} \\
&\textrm{Orbit\;III}_3: \qquad 
\color{CFTIII}{L(-\tfrac14,\,\tfrac{29}{56})\;\xrightarrow{\psi}\;L(0,\,\tfrac{1}{7})\;\xrightarrow{\psi}\;L(-\tfrac34,\,\tfrac{29}{56})\;\xrightarrow{\psi}\;L(-\tfrac12,\,\tfrac{9}{14})\;\xrightarrow{\psi}\;L(-\tfrac14,\,\tfrac{29}{56})} 
    \label{eq:sf5}
\end{align}
Besides the vacuum orbit III$_1$ and the lowest-weight orbit III$_2$ there is another orbit III$_3$. A distinguishing feature of CFT III is that all three orbits of spectral flow have the same length of four.

\paragraph{\color{CFTIV}{CFT IV.}} Besides the vacuum orbit starting from $L(0,\,0)$ and the lowest-weight orbit starting from $L(-\tfrac16,\,\tfrac{1}{12})$ --- again we use \eqref{eq:sf2} to determine $h$ from $q=-\tfrac16$ and $k=\tfrac23$ --- there are two additional orbits of spectral flow. For both of them we use a starting module with $q=-\tfrac13$, in one case with $i=1$ using \eqref{eq:sf2} and in the other case with $i=2$ using \eqref{eq:sf4}. The four spectral flow orbits of CFT IV are given by
\begin{align}
&\textrm{Orbit\;IV}_1: \quad 
\color{CFTIV}{L(0,\,0)\;\xrightarrow{\psi}\;L(-\tfrac23,\,\tfrac13)\;\xrightarrow{\psi}\;L(-\tfrac43,\,\tfrac43)\;\xrightarrow{\psi}\;L(0,\,1)\;\xrightarrow{\psi}\;L(-\tfrac23,\,\tfrac43)\xrightarrow{\psi}\;L(\tfrac23,\,\tfrac13)\;\xrightarrow{\psi}\;L(0,\,0)} \\
&\textrm{Orbit\;IV}_2: \quad 
\color{CFTIV}{L(-\tfrac16,\tfrac{1}{12})\,\xrightarrow{\psi}\,L(-\tfrac{5}{6},\tfrac{7}{12})\,\xrightarrow{\psi}\,L(-\tfrac12,\tfrac{3}{4})\,\xrightarrow{\psi}\,L(-\tfrac16,\tfrac{7}{12})\,\xrightarrow{\psi}\,L(\tfrac16,\tfrac{1}{12})\,\xrightarrow{\psi}\,L(-\tfrac12,\tfrac{1}{4})\,\xrightarrow{\psi}\,L(-\tfrac16,\tfrac{1}{12})} \\
&\textrm{Orbit\;IV}_3: \quad 
\color{CFTIV}{L(-\tfrac13,\,\tfrac{2}{15})\;\xrightarrow{\psi}\;L(-1,\,\tfrac{4}{5})\;\xrightarrow{\psi}\;L(\tfrac13,\,\tfrac{2}{15})\;\xrightarrow{\psi}\;L(-\tfrac13,\,\tfrac{2}{15})} \\
&\textrm{Orbit\;IV}_4: \quad 
\color{CFTIV}{L(-\tfrac13,\,\tfrac{8}{15})\;\xrightarrow{\psi}\;L(0,\,\tfrac{1}{5})\;\xrightarrow{\psi}\;L(-\tfrac23,\,\tfrac{8}{15})\;\xrightarrow{\psi}\;L(-\tfrac13,\,\tfrac{8}{15})} 
    \label{eq:sf6}
\end{align}
Besides the vacuum orbit IV$_1$ and the lowest-weight orbit IV$_2$ there are two additional orbits IV$_3$ and IV$_4$. A distinguishing feature of CFT IV is that the former (latter) two orbits have length six (three).

\subsection{Chern--Simons formulation of conformal gravity}\label{app:F}

Conformal gravity in three dimensions \cite{Deser:1982vy} can be formulated as a CS theory \cite{Horne:1988jf}. The bulk action
\begin{equation}
I[A] = \frac{k_{\textrm{\tiny CS}}}{4\pi}\,\int \Big< A\wedge\extd A+\frac23\,A\wedge A\wedge A\Big>
\end{equation}
has the CS level $k_{\textrm{\tiny CS}}$ as the only coupling constant. The non-abelian connection $A$ has the gauge algebra $\mathfrak{so}(3,2)$, i.e., the conformal algebra in three spacetime dimensions with signature $(-,\,+,\,+)$. The bilinear form $\langle\,,\,\rangle$ is determined uniquely up to normalization \cite{Horne:1988jf}. 

The gauge algebra $\mathfrak{so}(3,2)$ in a convenient basis is given by the commutation relations (commutators not displayed vanish; $a,b\in\{-1,\,0,\,1\}$)
\begin{align}
[L_a,\,L_b] &= (a-b)\,L_{a+b} & [L_a,\,M_b] &= (a-b)\,M_{a+b} &  [L_a,\,K_b] &= (a-b)\,K_{a+b} \\
[J_0,\,M_a] &= M_a & [J_0,\,K_a] &= -K_a & [M_a,\,K_b] &= (a-b)\,L_{a+b}-\big(a^2+b^2-ab-1\big)\,J_0 \,.
\end{align}
The algebra above is the linear classical wedge algebra associated with the quantum conformal BMS\(_3\) algebra. It is obtained by restricting the integer indices $n$ to the set $a\in\{-1,\,0,1\}$ for $L_n$, $M_n$ and $K_n$ and to $0$ for $J_n$. Note that all anomalous terms drop out due to these restrictions of the index sets. Thus, all the ``super-'' prefixes can be dropped and the algebra above generates the usual (Lorentz-)rotations $L_a$, translations $M_a$, SCTs $K_a$ and dilatations $J_0$. Moreover, in the commutator $[M_a,\,K_b]$ all non-linear terms are dropped in the wedge algebra and the semiclassical relation $c=-12k$ is inserted. 

For calculations it can be useful to have a matrix representation of the gauge algebra in the basis above. Due to the well-known isomorphism of the semisimple Lie algebras $B_2\simeq C_2$ we can exploit $\mathfrak{so}(3,2)\simeq\mathfrak{sp}(4)$ and use $4\times 4$ instead of $5\times 5$ real matrices (for an explicit realization in terms of the latter see Eq.~(6) in \cite{Horne:1988jf}).
\begin{align}
L_0 &= \frac12\,\begin{pmatrix}
-1&0&0&0 \\
0&1&0&0 \\
0&0&1&0 \\
0&0&0&-1
\end{pmatrix} &
L_+ &= \begin{pmatrix}
0&1&0&0 \\
0&0&0&0 \\
0&0&0&0 \\
0&0&-1&0
\end{pmatrix} &
L_- &= \begin{pmatrix}
0&0&0&0 \\
-1&0&0&0 \\
0&0&0&1 \\
0&0&0&0
\end{pmatrix} \\
M_0 &= \frac{1}{\sqrt{2}}\,\begin{pmatrix}
0&0&0&1 \\
0&0&1&0 \\
0&0&0&0 \\
0&0&0&0
\end{pmatrix} &
M_+ &= \sqrt{2}\,\begin{pmatrix}
0&0&1&0 \\
0&0&0&0 \\
0&0&0&0 \\
0&0&0&0
\end{pmatrix} &
M_- &= \sqrt{2}\,\begin{pmatrix}
0&0&0&0 \\
0&0&0&1 \\
0&0&0&0 \\
0&0&0&0
\end{pmatrix} \\
K_0 &= \frac{1}{\sqrt{2}}\,\begin{pmatrix}
0&0&0&0 \\
0&0&0&0 \\
0&1&0&0 \\
1&0&0&0
\end{pmatrix} &
K_+ &= \sqrt{2}\,\begin{pmatrix}
0&0&0&0 \\
0&0&0&0 \\
0&0&0&0 \\
0&-1&0&0
\end{pmatrix} &
K_- &= \sqrt{2}\,\begin{pmatrix}
0&0&0&0 \\
0&0&0&0 \\
-1&0&0&0 \\
0&0&0&0
\end{pmatrix} \\
J_0 &= \frac12\,\begin{pmatrix}
1&0&0&0 \\
0&1&0&0 \\
0&0&-1&0 \\
0&0&0&-1
\end{pmatrix} &&&&
\end{align}

In the basis above, the bilinear form $\langle G_1,\,G_2\rangle=\tr(G_1\,G_2)$ is given by 
\begin{align}
\langle L_0,\,L_0\rangle &= 1 & 
\langle L_+,\,L_-\rangle &= -2 &
\langle M_0,\,K_0\rangle &= 1 &
\langle M_\pm,\,K_\mp\rangle &= -2 &
\langle J_0,\,J_0\rangle &= 1
\end{align}
where combinations not displayed vanish or are determined by symmetry, $\langle G_1,\,G_2\rangle=\langle G_2,\,G_1\rangle$.

Fuentealba et al.~imposed suitable boundary conditions and obtained a semiclassical limit of $W_{(2,2,2,1)}$ as asymptotic symmetry algebra \cite{Fuentealba:2020zkf}. Apart from changing commutators to Poisson brackets (with the usual factor $i$), the algebraic relations \eqref{eq:1}-\eqref{eq:8} are unchanged but Eq.~\eqref{eq:angelinajolie} reduces to $c=-12k$ and the nonlinear relation \eqref{eq:9} changes to
\begin{multline}
i\,\big\{M_n,\,K_m\big\} = \frac{c}{12}\,\big(n^3-n\big)\,\delta_{n+m,\,0} + (n-m)\,L_{n+m} - \big(n^2+m^2-nm-1\big)\,J_{n+m} \\
- \frac{2}{k}\,(n-m)\,\sum_r J_{n+m-r}\,J_r + \frac{2}{k}\,\sum_rL_{n+m-r}\,J_r - \frac{2}{k^2}\,\sum_{r,s}J_{n+m-r-s}\,J_r\,J_s\,.
\end{multline}

\end{document}